\documentclass[11pt]{article}
\usepackage{graphicx}

\title{Exotic mesons from quantum chromodynamics with improved gluon and quark actions on the anisotropic lattice}
\vspace{2cm}
\author{Zhong-Hao Mei, 
and Xiang-Qian Luo\thanks{Corresponding author. Email: stslxq@zsu.edu.cn}\\
{\small\sl Department of Physics, Zhongshan (Sun Yat-Sen) University, Guangzhou 510275, China}\\
}

\date{\today}

\begin{document}
\maketitle

\begin{abstract}
Hybrid (exotic) mesons, which are important predictions of quantum chromodynamics (QCD), 
are states of quarks and anti-quarks bound by excited gluons.
First principle lattice study of such states 
would help us understand the role of ``dynamical'' color in low energy QCD
and provide valuable information for experimental search for these new particles. 
In this paper, we apply both improved gluon and quark actions to the hybrid mesons,
which might be much more efficient than the previous works
in reducing lattice spacing error and finite volume effect. 
Quenched simulations were done at $\beta=2.6$ and on a $\xi=3$ anisotropic $12^3\times36$ lattice using our PC cluster.
We obtain $2013 \pm 26 \pm 71$ MeV for the mass of the $1^{-+}$ hybrid meson ${\bar q}qg$ in the light quark sector, 
and  $4369 \pm 37 \pm 99$Mev in the charm quark sector;
the mass splitting between the $1^{-+}$ hybrid meson ${\bar c}c g$ in the charm quark sector and the 
spin averaged S-wave charmonium mass is estimated to be $1302 \pm 37 \pm 99$ MeV. As a byproduct,
we obtain  $1438 \pm 32 \pm 57$ MeV for  the mass of a P-wave $1^{++}$ ${\bar u}u$ or ${\bar d}d$  meson
and $1499  \pm 28 \pm 65$ MeV for the mass of a  P-wave $1^{++}$ ${\bar s}s$ meson,
which are comparable to their experimental value  1426 MeV for the $f_1(1420)$ meson. 
The first error is statistical, and the second one is systematical.
The mixing of the hybrid meson with a four quark state is also discussed.
\end{abstract}


\leftline{{\bf PACS } number: 12.39.Mk, 12.38.Gc, 11.15Ha}
\leftline{Published in {\it International Journal of Modern Physics} {\bf A18} (2003) 5713 -5724.}
\setcounter{page}{0}
\newpage

\section{Introduction}

The hybrid mesons  ${\bar q} qg$, which are bound states of quark $q$, anti-quark ${\bar q}$ and excited gluons $g$,
have exotic quantum numbers inaccessible to $\bar{q}q$ mesons.
There is increasing interest by BES and CLEO collaborations 
as well as experimentalists at Brookhaven and Jefferson labs in searching these new particles.
Monte Carlo lattice technique is the ideal method not only for computing $\bar{q}q$ meson spectrum, 
but also for hybrids. Of course, the lattice QCD approach is not free of systematical errors.
As is well known, the Wilson gauge and quark actions\cite{Wilson} 
have $O(a^2)$ and  $O(a)$  errors respectively,
where $a$ is the lattice spacing. These errors are smaller only at very small bare coupling,
and very large lattice volume is required to get rid of finite size effects.

The idea of Symanzik improvement\cite{Symanzik:1983dc} is to reduce the lattice spacing errors,
by adding new terms to the Wilson actions.
Together with the tadpole\cite{Lepage:1992xa} or non-perturbative improvement\cite{Luscher:1998pe},
the pursuit of the Symanzik program has recently led to significant progress in lattice
gauge theory, opening the possibility to approaching continuum physics on coarse and small lattices. 
Simulations on anisotropic lattices help getting very good signal in spectrum computations.

There have been several quenched lattice calculations of hybrid meson masses, 
in either the light quark or heavy quark sector.
In Refs.\cite{Griffiths:1983ah,Perantonis:1990dy,Bernard:1997ib}, 
the Wilson gluon action and quark action were used.
In Refs.\cite{Lacock:1996ny,McNeile:1998cp}, 
the authors used Wilson gauge action and SW (clover) improved quark action\cite{Sheikholeslami:1985ij}. 
For the hybrid mesons containing heavy quarks ${\bar Q}Q$g, 
Non-relativistic QCD (NRQCD) quark action\cite{Manke:1998qc,Drummond:1999db} 
and LBO (leading Born-Oppenheimer) quark action\cite{Juge:1999ie}
were also applied to the study of charmonium. 
As mentioned in Refs.\cite{Klassen:1998fh,Chen:2000ej,Okamoto:2001jb},
these approaches have advantages and disadvantages.

In this work, {\it both improved gluon and quark actions on
the anisotropic lattice} are employed, which should have smaller systematical errors, 
and should be more efficient in reducing 
the lattice spacing and finite volume effects. We will present
results for the $1^{-+}$ hybrid mass
and the splitting between the $1^{-+}$ hybrid mass and the 
spin averaged S-wave mass for charmonium. As a byproduct, we will also estimate the $f_1(1420)$ mass.

The remaining part of the paper is organized as follows. 
We describe the improved gluon and quark actions in Sec.\ref{secII},
and operators for the ${\bar q}q$ mesons and ${\bar q} q g$ mesons in Sec.\ref{secIII}.
Simulation details are given in Sec.\ref{secIV} and results
and conclusions are summarized in Sec.\ref{secV}.

\section{Improved actions on the anisotropic lattice}
\label{secII}

The total lattice action is $S=S_g+S_q$. 
$S_g$ is the classical-improved gluon action\cite{Morningstar:1997ff,Luo:1998dx,Alford:2000an}:
\begin{eqnarray}
    S_g=-{6 \over g^2} {1 \over \xi_0} \sum_{x,j<k}({5 \over 3}P_{jk} 
-
{1 \over 12} R_{jk}- {1 \over 12} R_{kj})
- {6 \over g^2} \xi_0 \sum_{x,j}({4 \over 3} P_{j4}
-
{1 \over 12} R_{j4}),
\label{classical}
\end{eqnarray}
which differs from the continuum Yang-Mills action in imaginary time 
$\int d^4 x ~ F^{\alpha}_{\mu \nu} F^{\alpha}_{\mu \nu}/4$
by $O(a_t^2)$ and $O(a_s^4)$ at classical level, with $a_t$ and $a_s$
being the temporal and spatial lattice spacing respectively. 
$g$ is the gauge coupling; 
$x$ is the lattice site and the subindeces   $j, k$ and $4$ in Eq.(\ref{classical}) are respectively 
the spatial and temporal
directions; $\xi_0$ is the bare anisotropic parameter.
$P$ stands for a $1\times 1$ plaquette and $R$ for  a $2\times 1$ rectangle. More explicitly,
they are
\begin{eqnarray}
P_{jk} &=&
\frac{1}{3} {\rm Tr ~ Re} \left[ U_{j}(x)U_k(x+{\hat j}) U^{\dagger}_{j} (x+ {\hat k}) U^{\dagger}_{k}(x)\right],
\nonumber \\
P_{i4} &=& 
\frac{1}{3} {\rm Tr ~ Re} \left[  U_{j}(x)U_4(x+{\hat j}) U^{\dagger}_{j} (x+ {\hat 4}) U^{\dagger}_{4}(x)  \right],
\nonumber \\
R_{jk} 
&=&
{1 \over 3} {\rm Re ~ Tr} \left[U_j(x) U_j(x+\hat{j}) U_k(x+2\hat{j}) 
U^{\dagger}_j(x+\hat{k}+\hat{j})
U^{\dagger}_j(x+\hat{k}) U^{\dagger}_k(x) \right],
\nonumber \\
R_{j4} 
&=& 
{1 \over 3} {\rm Re ~ Tr} \left[ U_j(x) U_j(x+\hat{j}) U_4(x+2\hat{j}) 
U^{\dagger}_j(x+\hat{j}+\hat{4}) 
U^{\dagger}_j(x+\hat{4}) U^{\dagger}_4(x) \right],
\end{eqnarray}
where $U_j(x)$ is the gauge link variable at site $x$ and $j$-th direction.
The clover action for quarks
\cite{Klassen:1998fh,Chen:2000ej,Okamoto:2001jb} is 
\begin{eqnarray}
    S_q &=& \sum_{x} {\bar \psi} (x)\psi(x)- \kappa_t \sum_{x}
[\bar{\psi}(x) (1-\gamma_4)U_{4}(x)
\psi(x+\hat{4}) + {\bar \psi}(x)(1+\gamma_4)
U^{\dag}_{4}(x) \psi (x-\hat{4})]
\nonumber \\
&-& \kappa_s \sum_{x,j} [{\bar \psi}(x) (r_s-\gamma_j)U_{j}(x) \psi (x+\hat{j})
+ {\bar \psi} (x) (r_s+\gamma_j)U^{\dag}_{j}(x-\hat{j})\psi(x-\hat{j})]
\nonumber \\
&+& i {\kappa}_s {C}_s  \sum_{x,j<k} {\bar \psi} (x)
\sigma_{jk}{\hat F_{jk}}(x)\psi (x) + i \kappa_s C_t \sum_{x,j}{\bar \psi} (x)
\sigma_{j4} {\hat F_{j4}} (x) \psi (x),
\label{S_q}
\end{eqnarray}
which differs at classical level from the continuum Dirac action 
$\int d^4 x ~ {\bar q}(\gamma_{\mu}D_{\mu}+m_q)q$ by $O(a^2)$.
(There is another way to reduce the lattice spacing error using the Hamber-Wu quark action\cite{Hamber:1983qa}.) 
The continuum quark field $q$ is related to the lattice counter part $\psi$ by $\psi=\sqrt{a_s^3/(2\kappa_t)}q$,
and the bare quark mass is related to $\kappa_t$ and $\kappa_s$ by 
\begin{eqnarray}
a_tm_q \equiv 1/(2\kappa_t)-3r_s \kappa_s/\kappa_t-1.
\end{eqnarray}
For the gauge field tensor ${\hat F}$ we use the clover-leaf construction\cite{Sheikholeslami:1985ij,Luo:1996tx}, and
the clover coefficients at classical level are
\begin{eqnarray}
C_s=r_s, ~~~ C_t \approx \frac{1+ \xi_0 r_s}{2},
\end{eqnarray}
where $r_s$ is the Wilson parameter and it is usually taken to be 1.

The tadpole-improved actions are more continuum-like, and can be obtained by following replacement
\begin{eqnarray}
U_j(x) \to {U_j (x) \over u_s}, ~~~ U_4(x) \to {U_4 (x) \over u_t},
\nonumber \\
g^2 \to {\tilde g}^2 ={g^2 \over u_s^3 u_t}, ~~~ \xi_0 \to \xi={u_t \over u_s} \xi_0,
\nonumber \\
\kappa_s \to {\tilde \kappa_s}=u_s\kappa_s, ~~~ \kappa_t \to {\tilde \kappa_t}=u_t\kappa_t,
\nonumber \\
C_s \to {\tilde C_s}=u^3_s C_s, ~~~ C_t \to {\tilde C_t} \approx u_s u^2_t {1+\xi r_s \over 2}.
\end{eqnarray}
$\xi$ is just the aspect ratio of spatial and temporal lattice spacings $a_s/a_t$;
$u_s$ and $u_t$ are tadpole parameters are taken to be
\begin{eqnarray}
u_s=\langle P_{jk} \rangle^{1/4}, ~~~ u_t=1.
\end{eqnarray}

\section{Meson operators and green functions}
\label{secIII}

The hybrid operators are constructed by combining a quark, an anti-quark and 
the chromo-electric or chromo-magnetic
field to form a color singlet with the given spin $J$, parity $P$ and charge conjugation $C$.
The generic structure is ${\bar \psi^{c_1}}_{f_1}\Gamma\psi^{c_2}_{f_2}{\hat F}^{c_1 c_2}$, 
where $c_1$ and $c_2$ 
stand for color indices, and $f_1$ and $f_2$ for flavor indices; 
$\Gamma$ is some combination of Dirac matrices and spatial derivatives. 
The details are given in Ref.\cite{Bernard:1997ib}. For convenience, 
we present in Tab. \ref{tab1} various source and sink operators 
for ${\bar q}q$ mesons, ${\bar q}qg$ mesons and a four quark state 
we have used. We consider such a $q^4$ operator (four quark state) because it can mix with the hybrid mesons
through the hairpin diagram.

\begin{table} 
  \begin{center}
  \begin{tabular}{|c|c|c|c|c|}\hline
 Light meson        & Charmonium   &  $J^{PC}$ &  Mnemonic      & Operator                    \\ \hline
    ${\bar q} q$    & ${\bar c}c$  &           &                &                             \\
   $\pi$            &  $\eta_c$    &  $0^{-+}$ & $^1S_0 ~\pi$   & ${\bar \psi}\gamma_{5}\psi$  \\
   $\rho$           &  $J/\psi$    &  $1^{--}$ & $^3S_1 ~ \rho$ & ${\bar \psi}\gamma_{i}\psi$  \\
$f_1({\rm P-wave})$ &  $\chi_{c_1}$ & $1^{++}$  & $^3P_1 ~ f_1$  & $\epsilon_{ijk}{\bar \psi}\gamma_{j}\stackrel{\leftrightarrow}{\partial_{k}}\psi$    \\ \hline

   ${\bar q} q g$    &  ${\bar c}cg$ &           &                &                               \\
   $1^{-+}$         & $1^{-+}$      & $1^{-+}$  & $\rho\bigotimes B$ & ${\bar \psi}^{c_1}\gamma_{j}\psi^{c_2}F^{c_1c_2}_{ji}$   \\
   $q^{4}$          &    $Q^{4}$    &   $1^{-+}$ & $\pi \bigotimes a_{1}$  & ${\bar \psi}^{c_1}_{f_1}(\vec{x})\gamma_{5}
      \psi_{f_2}^{c_1}(\vec{x}){\bar \psi}^{c_2}_{f_2}(\vec{y})\gamma_{5}\gamma_{i}\psi^{c_2}_{f_3}(\vec{y})$  \\ 
\hline       
\end{tabular}
\end{center}
\caption{\label{tab1} Source and sink operators used in Eq.(\ref{correlation}), 
the two point Green function of mesons.}
\end{table}

In Tab.\ref{tab1}, 
the chromo-electric and chromo-magnetic fields bring in $J^{PC}=1^{--}$ and $1^{+-}$ respectively,
$\epsilon_{ijk}\bar{\psi}\gamma_{j}\stackrel{\leftrightarrow}{\partial_{k}}\psi$
is a P-wave operator for $f_{1}$ with $J^{PC}=1^{++}$. The
operator $\stackrel{\leftrightarrow}{\partial_{i}}=\stackrel{\rightarrow}{\partial_{i}}-\stackrel{\leftarrow}{\partial_{i}}$ 
have  $J^{PC}=1^{--}$,  where $C=-1$ is because 
that $C$ interchanges the quark and the anti-quark.

To use information from the different source and sink operators, we fit 
simultaneously the correlation function of two or more operators with different amplitudes $A_i$ and $A_j$ for each
source $O_i$ and sink $O_j$ listed in Tab.\ref{tab1}, but with the same masses $m_{\rm eff}$ for all of them. 
For large $t$, it behaves as
\begin{eqnarray}
   \sum_{\vec{x}} <O_{i}(\vec{0},0)O_{j}(\vec{x},t)>=A_{i}A_{j}e^{-m_{\rm eff} a_t t} +...
\label{correlation}
\end{eqnarray}

\section{Simulation results}
\label{secIV}

Part of our work is based on the MILC public code\cite{s18}, which 
was written for isotropic lattices. We modified the code to anisotropic lattices.
To check the new code for clover quarks,  we computed some charmonium masses 
using the gauge configurations generated by the standard one-plaquette gauge action, 
and compared with those of the  CP-PACS collaboration\cite{Okamoto:2001jb}.
To check the new code for gluons with the tadpole improved gauge action,
we calculated the tadpole parameter $u_s$ 
and compared results with those of  Alford et al.\cite{Alford:2000an}.
The simulations were performed 
on our PC cluster for parallel computing\cite{s12,Luo:2001mp,Luo:2001yt,Luo:2002kr,Mei:2002fu}. 
The results are well consistent with the literature, and details of comparison will be reported elsewhere.

The results presented in this paper use the SU(3) pure gauge configurations 
generated with the tadpole-improved gluon action  
\begin{eqnarray}
    S_g=-\beta {1 \over \xi} \sum_{x,j<k} \left( {5 \over 3} {P_{j,k} \over u_s^4}
-{1 \over 12} {R_{j,k} \over u_s^6} - {1 \over 12} {R_{k,j} \over u_s^6} \right)
- \beta \xi \sum_{x,j} \left( {4 \over 3} {P_{j,4} \over u_s^2}
- {1 \over 12} {R_{j,4} \over u_s^4} \right),
\end{eqnarray}
using Cabibbo-Marinari pseudo-heatbath algorithm. The configurations are decorrelated by 
SU(2) sub-group over-relaxations. We calculated the
tadpole parameter $u_s$ in a self-consistent manner. 
At $\beta=2.6$ and $\xi=3$, we generated 90 pure gauge configurations on the $12^3 \times 36$ lattice. 
Although such an ensemble is not very big, it is bigger than earlier 
simulations by UKQCD 
and MILC collaborations\cite{Perantonis:1990dy,Bernard:1997ib,Lacock:1996ny,McNeile:1998cp} on isotropic lattices. 
To practically implement the tadpole-improvement for the quarks, it is more convenient to
tune the parameters $\kappa_t$, $\kappa_s$ (or $\zeta=\kappa_t/\kappa_s$), 
$C_t^{\rm TI}$ and $C_s^{\rm TI}$ of the following quark action 
\begin{eqnarray}
    S_q &=& \sum_{x} {\bar \psi} (x)\psi(x)- \kappa_t \sum_{x}
[\bar{\psi}(x) (1-\gamma_4)U_{4}(x)
\psi(x+\hat{4}) + {\bar \psi}(x)(1+\gamma_4)
U^{\dag}_{4}(x) \psi (x-\hat{4})]
\nonumber \\
&-& \kappa_s \sum_{x,j} [{\bar \psi}(x) (1-\gamma_j)U_{j}(x) \psi (x+\hat{j})
+ {\bar \psi} (x) (1+\gamma_j)U^{\dag}_{j}(x-\hat{j})\psi(x-\hat{j})]
\nonumber \\
&+& i \kappa_s C_s^{\rm TI}  \sum_{x,j<k} {\bar \psi} (x)
\sigma_{jk}{\hat F_{jk}}(x)\psi (x) + i \kappa_s C_t^{\rm TI} \sum_{x,j}{\bar \psi} (x)
\sigma_{j4} {\hat F_{j4}} (x) \psi (x),
\end{eqnarray}
where the clover constants are
\begin{eqnarray}
C^{\rm TI}_s=\frac{1}{u_s^3}, ~~~ C^{\rm TI}_t=\frac{1}{2}(1+\frac{1}{\xi})\frac{1}{u_s^2},
\end{eqnarray}
without rescaling the link variables.
The simulation parameters are listed in Tab.\ref{tab3}.

\begin{table}
  \begin{center}
  \begin{tabular}{|c|c|c|c|c|c|c|c|}\hline
   $\beta$ & $\xi$  & $u_s$  & $\kappa_t$                     & $\zeta$ & $C_t^{\rm TI}$ & $C_s^{\rm TI}$& $L^3\times T$ 
\\ \hline
   2.6     & 3      & 0.8192 & 0.4119, 0.4199, 0.4279, 0.4359 &   3     & 2.441 & 1.818 & $12^3\times 36$
      \\ \hline               \end{tabular}
\end{center}
\caption{\label{tab3} Simulation parameters for hybrid meson spectrum calculations.}
\end{table}

We obtained the quark propagator by inverting the quark matrix $\Delta$ 
in the tadpole-improved $S_q=\sum_{x,y} {\bar \psi}(x) \Delta_{x,y} \psi (y)$  
by means of BICGStab algorithm. The residue is of $O(10^{-7})$.
We evaluated the correlation functions with
various sources and sinks listed in Tab. \ref{tab1}, at four values of the Wilson hopping parameter
($\kappa_t=0.4119$, 0.4199, 0.4279, 0.4359).

\begin{figure}[hb]
\begin{center}
\rotatebox{270}{\includegraphics[width=8cm]{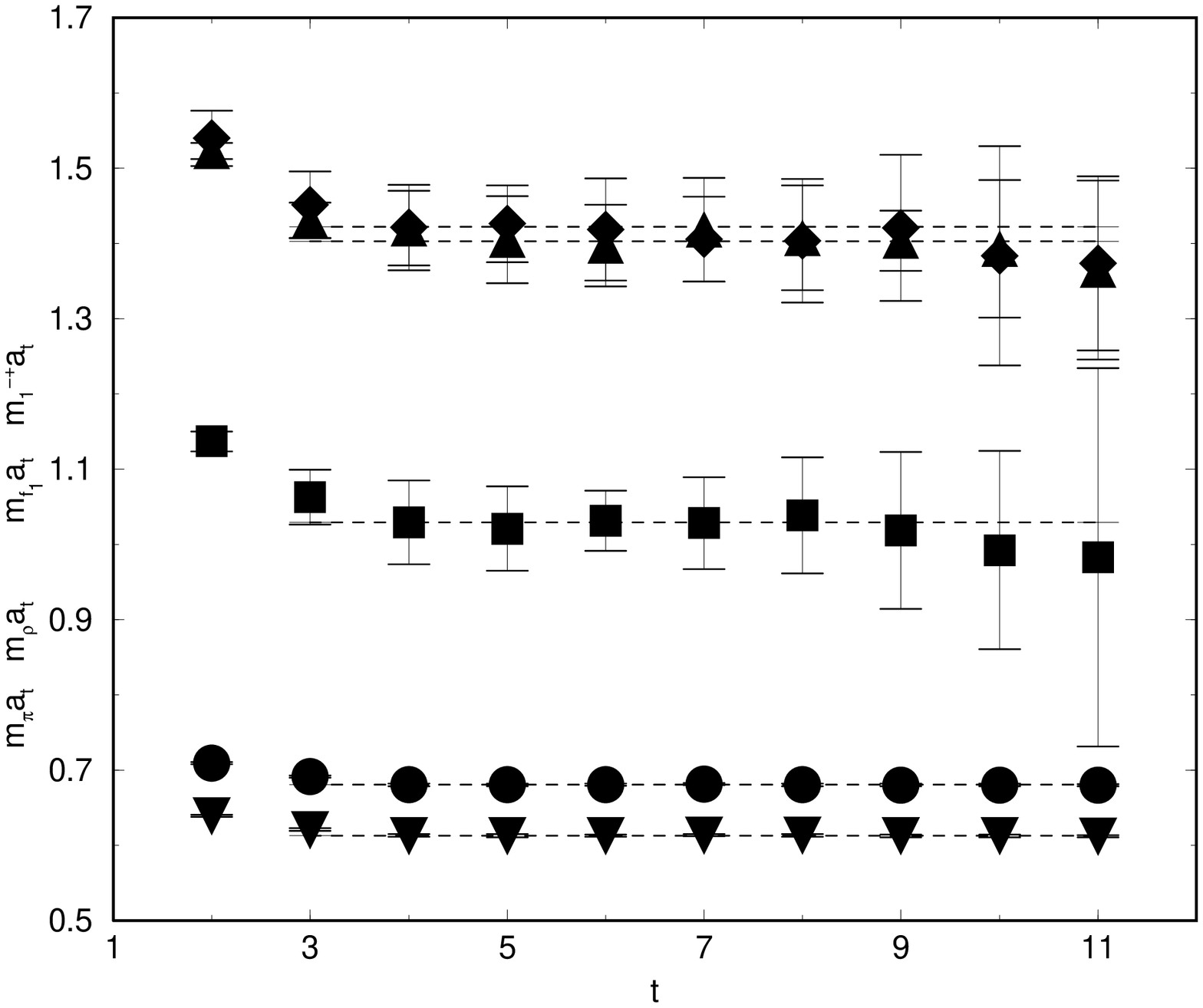}}
\end{center}
\caption{\label{mprf1t.eps}  The effective masses for the $\pi$ (triangle down),
 $\rho$ (circles), $f_1$ P-wave (square) mesons 
and $1^{-+}$  exotic meson (the diamond for $1^{-+}$ source and the triangle
up for the $q^4$ source) at $\kappa_t=0.4359$.
The error bars reflect 
the sum of statistical and systematic errors.
  }
\end{figure}

The effective mass is obtained by fitting the correlation function to Eq.(\ref{correlation}). 
In Fig.\ref{mprf1t.eps}, we show the effective masses for the $\pi$, $\rho$, $f_1$ ordinary mesons and 
$1^{-+}$ exotic meson at $\kappa_t=0.4359$. For the ordinary mesons, we used the their corresponding operators
in Tab. \ref{tab1} as both source and sink. For the exotic meson, we tried two different cases:
(1) the $1^{-+}$ operator as both source and sink; (2) the $q^4$ source and $1^{-+}$ sink, which give consistent results within error bars.

\begin{figure}[hb]
\begin{center}
\rotatebox{270}{\includegraphics[width=8cm]{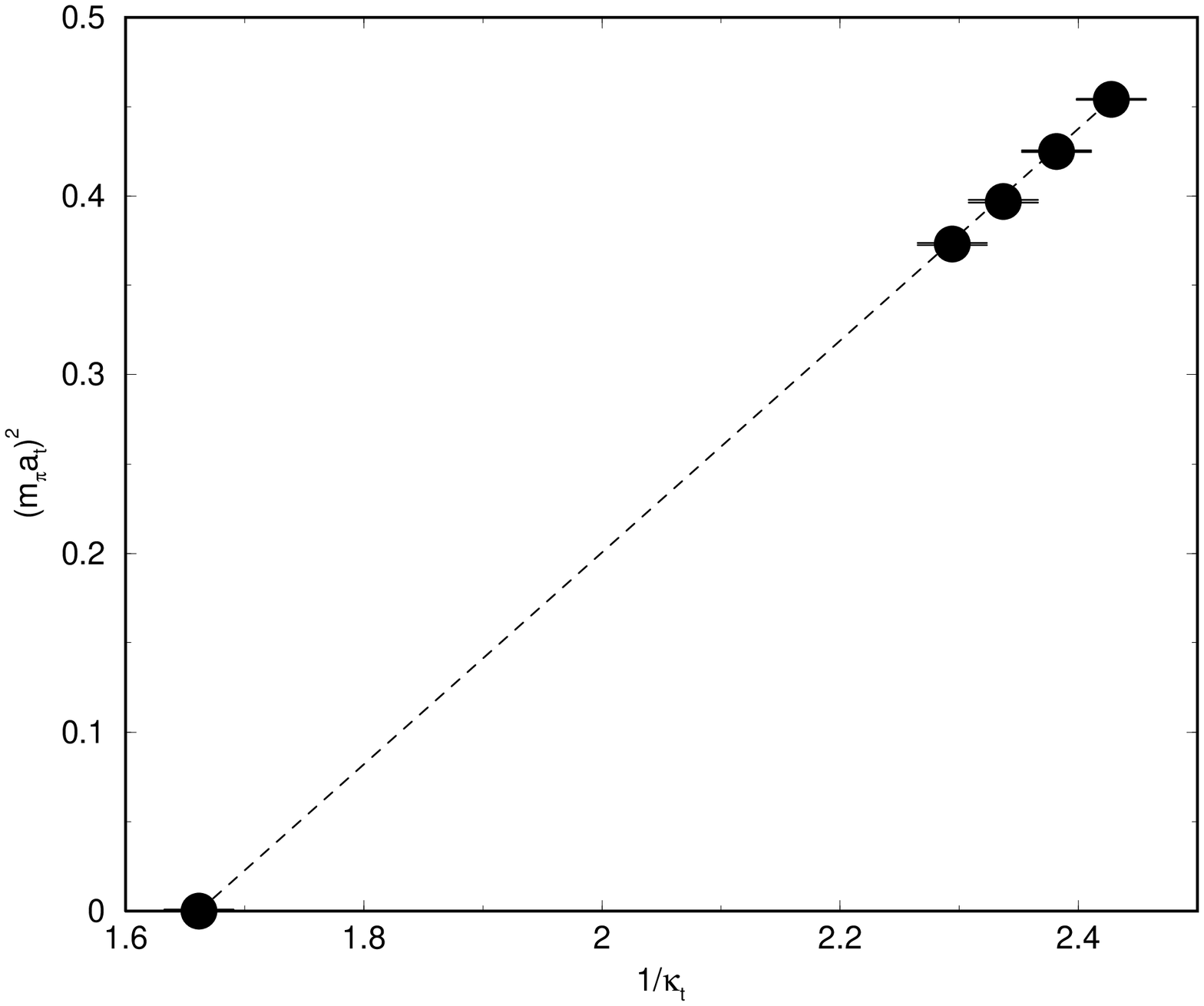}}
\end{center}
\caption{\label{mp2kat} Pion mass squared as a 
function of $1/\kappa_t$.
The error stands for  
the sum of statistical and systematic errors.}
\end{figure}

\begin{figure}[hb]
\begin{center}
\rotatebox{270}{\includegraphics[width=8cm]{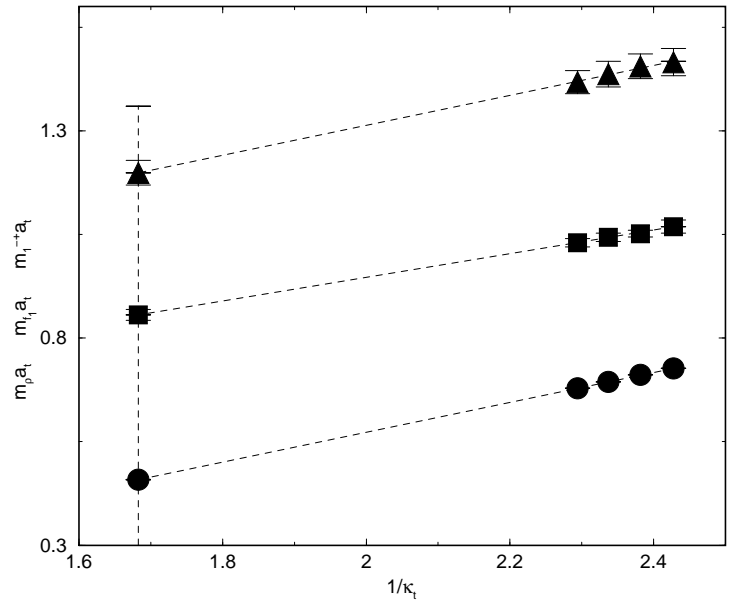}}
\end{center}
\caption{\label{mrf1kt.eps} Effective masses as a function of $1/\kappa_t$ 
and extrapolations to the chiral limit.  
The circles are $\rho$ masses for both
 $\rho$ source and  sink,
the squares are masses for the $f_{1}$ P-wave operator as both source and sink,
and the triangles are masses for the $1^{-+}\rightarrow 1^{-+}$ hybrid operator.
  }
\end{figure}

In Tab.\ref{tab4}, we present the data for ${\bar q}q$  meson masses at four different $\kappa_t$,
and in Tab.\ref{tab5} the data for the $1^{-+}$ exotic meson.
Fig.\ref{mp2kat} plots the  pion mass squared as a function of $1/\kappa_t$. 
By linearly extrapolating the data to the massless limit, we obtain $\kappa_t^{\rm chiral}=0.5945(3)(11)$, 
which is also shown in the figure.  
The regression functionality of the software $Xmgr$ is used in the extrapolation.
In Fig.\ref{mrf1kt.eps}, we show the effective mass of other mesons as a function of $1/\kappa_t$ 
and  extrapolation to the
chiral limit. 
In the chiral limit, we have $a_tm_{\rho}(\kappa_t \to \kappa_t^{\rm chiral})=0.458(3)(15)$. Using
the experimental value $m_{\rho}=768.5$MeV, we get $a_t^{-1}=1.679(2)(7)$GeV, and
\begin{eqnarray}
a_t \approx 0.11 {\rm Fermi}, ~~~ a_s \approx 0.33 {\rm Fermi}.
\end{eqnarray}
%


\begin{table} 
  \begin{center}
  \begin{tabular}{|c|c|c|c|c|}\hline
    Particle  & $ \kappa_t$  & Fit Range  &  $a_t m$ &  $ \chi^2/dof $  \\ \hline
    $\pi$     & 0.4119       &  8-16      &  0.674(3) &     5.1/7       \\
              & 0.4199       &  8-16      &  0.652(8) &     5.2/7       \\
              & 0.4279       &  8-16      &  0.630(8) &     3.5/7       \\
              & 0.4359       &  8-16      &  0.611(3) &     6.5/7        \\
              & $\kappa_t^{\rm chiral}$=0.5945     &  Extrapolation to the chiral limit   &  0 &    ---           \\ \hline              
\hline
    $\rho$    & 0.4119       &  8-16      &  0.727(7) &     3.5/7       \\
              & 0.4199       &  8-16      &  0.711(9) &     7.8/7       \\
              & 0.4279       &  8-16      &  0.694(8) &     6.6/7       \\
              & 0.4359       &  8-16      &  0.679(5) &     2.6/7        \\ 
              & $\kappa_t^{\rm chiral}$=0.5945     &  Extrapolation to the chiral limit   &  0.458(3)(15) &    ---  \\ \hline \hline
    $f_{1}$ P-wave   & 0.4119     &  3-11      &  1.069(17) &     4.9/5       \\
              & 0.4199     &  3-11      &  1.052(13) &     4.6/5       \\
              & 0.4279     &  3-11     &  1.043(14) &     8.9/5       \\
              & 0.4359     &  3-11      &  1.030(9) &     1.2/5        \\
              & $\kappa_t^{\rm chiral}$=0.5945     &  Extrapolation to the chiral limit   &  0.857(14)(27) &    ---         \\
 \hline
\end{tabular}
          \end{center}
\caption{\label{tab4} 
Effective masses of ordinary $\bar{q}q$ mesons at different $\kappa_t$. 
}
\end{table}
\begin{table}
  \begin{center}
  \begin{tabular}{|c|c|c|c|c|c|c|}\hline
  $\kappa_t$ & Source(s)$\rightarrow$Sink     & Fit Range & $ \chi^2/dof $ & $a_t m$    \\ \hline
   0.4119 & $1^{-+}\rightarrow 1^{-+}$      &  3-10      &   6.3/5     &   1.466(26)     \\
          & $q^4,1^{-+} \rightarrow 1^{-+}$ &  3-10     &   3.6/5     &   1.459(25)     \\ \hline
   0.4199 & $1^{-+}\rightarrow 1^{-+}$      &  3-10      &   6.8/5     &   1.456(28)     \\
          & $q^4,1^{-+} \rightarrow 1^{-+}$ &  3-10    &    5.5/5     &   1.449(36)     \\ \hline
   0.4279  & $1^{-+}\rightarrow 1^{-+}$      &  3-10      &    1.0/5     & 1.437(28)     \\
          & $q^4,1^{-+}\rightarrow 1^{-+}$   &  3-10      &    3.8/5     & 1.434(31)   \\ \hline
   0.4359  & $1^{-+}\rightarrow 1^{-+}$      &  3-10      &   3.3/5     &  1.418(35)     \\
          & $q^4,1^{-+}\rightarrow 1^{-+}$   &  3-10      &    7.0/5     & 1.400(37)    \\ \hline
$\kappa_t^{\rm chrial}=0.5945$ & $1^{-+}\rightarrow 1^{-+}$ & Extrapolation to the chiral limit & ---   & 1.197(8)(26) \\
$\kappa_t^{\rm charm}=0.1806$ & $1^{-+}\rightarrow 1^{-+}$ &  Tuning   & ---   & 2.603(12)(23) \\ \hline  
                         \end{tabular}
          \end{center}
\caption{\label{tab5} Effective mass of the exotic $1^{-+}$ meson at different $\kappa_t$.}
\end{table}

The UKQCD, MILC and CP-PACS collaborations used the unimproved Wilson gauge action to generate configurations.
They had to work on very large $\beta (> 6$), corresponding to very small $a_s (<$ 0.1 Fermi), to get rid of the
finite spacing errors.
They had also to use very large lattices $L^3 \ge 20^3$, to avoid strong lattice size effects at such small $a_s$.
In comparison, our lattices are much coarser, and the number of lattice sites is much smaller.
(Of course, finite size effects could be ignored because the physical size of our spatial lattice
$12^3 a_s^3=3.96^3 {\rm Fermi}^3$ should be big enough.)  
Our results for the effective mass indicate the existence of a very wide plateau,
which are better at least than the previous work on isotropic lattices.

\section{Predictions of masses and discussions}
\label{secV}

In this concluding section, we predict the $1^{-+}$ hybrid meson masses and
 discuss briefly the systematical errors.

By extrapolating the effective mass  of the $1^{-+}$ hybrid meson (Tab.\ref{tab5}) to the chiral limit, 
and using $a_t$ determined in 
Sec. \ref{secIV}, we get  $2013 \pm 26 \pm 71$ Mev. 
In Tab. \ref{tab6}, we compare the results from various lattice methods.
Our result is consistent with the data of MILC collaboration \cite{Bernard:1997ib}, 
who obtained using the Wilson gluon action and clover quark action
on  much larger isotropic lattices and much smaller $a_s$.

Also in  Tab. \ref{tab6}, we  present our results for the $1^{-+}$ hybrid meson mass in the charm quark sector,
using the method discussed in Refs. \cite{Bernard:1997ib,McNeile:1998cp}.
The corresponding $\kappa_t^{\rm charm}=0.1806(5)(18)$ 
is obtained by tuning $(m_{\pi}(\kappa_t \to \kappa_t^{\rm charm})+3m_{\rho} (\kappa_t \to \kappa_t^{\rm charm}))/4
=(m_{\eta_c}+3m_{J/\psi})/4$=3067.6MeV,
where on the right hand side, the experimental data $m_{\eta_c}= 2979.8$ Mev and $m_{J/\psi}=3096.9$ MeV are used.
The $1^{-+}$ hybrid meson mass at our
$1/\kappa_t^{\rm charm}$ is $m_{1^{-+}}= 4369 \pm 37 \pm 99$ MeV, 
which is consistent with the MILC data\cite{Bernard:1997ib}.
The splitting between the hybrid meson mass and the spin 
averaged S-wave mass [$m_{1^{-+}}-(m_{\eta_c}+3m_{J/\psi})/4$], at our $\kappa_t^{\rm charm}$ is
 $1302 \pm 37 \pm 99$ MeV, consistent with the CP-PACS data, obtained using the Wilson gluon action and NRQCD quark action
on much larger anisotropic lattices and much smaller $a_s$.

\begin{table}
  \begin{center}
  \begin{tabular}{|c|c|c|}\hline
 Light  $1^{-+} {\bar q}qg$  (GeV)     & Method &   Ref.  \\ \hline
   1.97(9)(30) & Isotropic $S_g({\rm W})+S_q({\rm W})$ & MILC97\cite{Bernard:1997ib}   \\
     1.87(20) & Isotropic $S_g^{\rm TI}({\rm W})+S_q^{\rm TI}({\rm SW})$  &UKQCD97\cite{Lacock:1996ny}    \\
   2.11(10) & Isotropic $S_g^{\rm TI}({\rm W})+S_q^{\rm TI}({\rm SW})$ & MILC99\cite{McNeile:1998cp} \\
  2.013(26)(71) & Anisotropic $S_g^{\rm TI}(1\times 1+ 2\times 1)+S_q^{\rm TI}({\rm SW})$  & {\bf ZSU (this work)} \\ \hline \hline
 $1^{-+} {\bar c}cg$  (GeV)      & Method &  Ref. \\ \hline
4.390 (80) (200) & Isotropic $S_g({\rm W})+S_q({\rm W})$ & MILC97\cite{Bernard:1997ib} \\
4.369 (37) (99)  & Anisotropic $S_g^{\rm TI}(1\times 1+ 2\times 1)+S_q^{\rm TI}({\rm SW})$ & {\bf ZSU (this work)} \\ \hline \hline
 $1^{-+} {\bar c}cg$ -1S ${\bar c}c$  splitting (GeV)     & Method &   Ref.  \\ \hline
   1.34(8)(20)& Isotropic $S_g({\rm W})+S_q({\rm W})$ & MILC97\cite{Bernard:1997ib}   \\
    1.22(15) & Isotropic $S_g^{\rm TI}({\rm W})+S_q^{\rm TI}({\rm SW})$ & MILC99\cite{McNeile:1998cp}   \\
    1.323(13) & Anisotropic $S_g^{\rm TI}({\rm W})+S_q^{\rm TI}({\rm NRQCD})$ & CP-PACS99\cite{Manke:1998qc}\\
  1.19 & Isotropic $S_g^{\rm TI}(1\times 1+ 2\times 1)+S_q^{\rm TI}({\rm LBO})$ & JKM99 \cite{Juge:1999ie}\\
   1.302(37)(99) & Anisotropic $S_g^{\rm TI}(1\times 1+ 2\times 1)+S_q^{\rm TI}({\rm SW})$ & {\bf ZSU (this work)} \\ \hline
                \end{tabular}
          \end{center}
\caption{\label{tab6} Predictions for the masses of $1^{-+}$ hybrid mesons ${\bar q}q g$ in the light quark sector 
and ${\bar c} c g$  in the  charm quark sector. 
Different lattice approaches are compared. Abbreviations: W for Wilson, $1\times 1+ 2\times 1$ for
the plaquette terms plus the rectangle terms,
SW for  Sheikholeslami-Wohlert (Clover), 
TI for tadpole-improved, NRQCD for non-relativistic QCD, and  LBO for leading Born-Oppenheimer.}
\end{table}

As a byproduct, in Tab. \ref{tab7}, we give the $f_1$ P-wave $1^{++}$ meson in the chiral limit,
as well as their experimental values\cite{Groom:in}.
If we assume that  the pion is massive and $f_1(1420)$ is made of ${\bar s}s$, 
the  $f_1$ P-wave $1^{++}$ meson mass would be $1499 \pm 28 \pm 65$ MeV.

\begin{table}
  \begin{center}
  \begin{tabular}{|c|c|c|c|}\hline
  $f_1$ P-wave (GeV)  & Method                                             &  Ref.  & Experiment \\ \hline
  1.438 (32) (57)    & Anisotropic $S_g^{\rm TI}(1\times 1+ 2\times 1)+S_q^{\rm TI}({\rm SW})$  
                                                         & {\bf ZSU (this work)} & 1.426 ~$f_1(1420)$? \\ \hline
                \end{tabular}
          \end{center}
\caption{\label{tab7} Predictions for the masses of the $1^{++}$ P-wave mesons ${\bar q}q$.
}
\end{table}

In conclusion, we have used the tadpole-improved gluon action and clover action to compute the
hybrid meson masses on much coarse anisotropic lattices. The main results are summarized in Tab.\ref{tab6} and
compared with other lattice approaches. 
In our opinion, 
our approach is much more efficient in reducing systematic errors due to finite lattice spacing and finite volume.

In our study, the quoted errors are statistical and systematical. We have not considered the quenching errors.
(Only in the case of unimproved Wilson gluon action plus the clover quark action, 
the effects of dynamical quarks have been studied\cite{Manke:2001ft} 
by applying the algorithm of Refs. \cite{Luo:1996tx,Aoki:2001pt},
and are found to have little influence on the spectrum of bottomonium spectrum.)
We have not taken into account the contamination of excited states when fitting the correlation function;
this is a common problem in lattice Lagrangian formulation;
Monte Carlo Hamiltonian method\cite{Jirari:1999jn,Luo:2001gb,Huang:1999fn}  proposed recently
might be helpful in this aspect.
Also, we have assumed that with the improved gluon and quark actions, 
the dependence of the results on the lattice spacings is negligible,
and we haven't extrapolated the results to the continuum limit. 
We hope to discuss these issues in the future.

\newpage

\noindent
{\bf Acknowledgements}

X.Q.L. is supported by the National Science Fund for Distinguished Young Scholars (19825117),
Key Project of National Science Foundation (10235040), 
Guangdong Natural Science Foundation (990212), 
National and Guangdong Ministries of Education, and
Hong Kong Foundation of the Zhongshan University 
Advanced Research Center.

This work was in part based on the MILC collaboration's public 
lattice gauge theory code. (See reference \cite{s18}.)  We are grateful to M. Alford, S. Aoki, 
C. DeTar, E. Gregory, U. Heller, C. Liu,  C. McNeile, C. Morningstar,
M. Okamoto, H. Shanahan, and D. Toussaint for helpful discussions.

\end{document}